\documentclass[11pt,twoside]{article}  
\usepackage{asp2008}
\usepackage{adassconf}

\begin{document}   

%
%

\paperID{A.11}

%

\title{Simulation and Fitting of Multi-Dimensional X-ray Data}

%
%
%
%
%

\markboth{DEWEY \& NOBLE}{Simulation and Fitting of Multi-Dimensional X-ray Data}

%
%
%
%

\author{Daniel Dewey, Michael S.\ Noble}
\affil{MIT Kavli Institute, Cambridge, MA, USA}

%

\contact{Dan Dewey}
\email{dd@space.mit.edu}

%
%
%

\paindex{Dewey, D.}
\aindex{Noble, M.S.}     

%



\begin{abstract}          
Astronomical data generally consists of 2 or more high-resolution axes,
e.g., X,Y position on the sky or wavelength and
position-along-one-axis (long-slit spectrometer). Analyzing these
multi-dimension observations requires combining 3D source models (including
velocity effects), instrument models, and multi-dimensional data comparison
and fitting. A prototype of such a "Beyond XSPEC" (Noble \& Nowak, 2008)
system is presented here using {\it Chandra} imaging and
dispersed HETG grating data. Techniques used include: Monte Carlo event
generation, chi-squared comparison, conjugate gradient fitting adapted to
the Monte Carlo characteristics, and informative visualizations at each
step. These simple baby steps of progress only scratch the surface of the
computational potential that is available these days for astronomical
analysis.
\end{abstract}

%
%

\section{Multi-Dimensional X-ray Data and ``Event-2D''}

Modern astronomical data often consists of 2 high-resolution dimensions:
e.g., \{$X,Y$\} for 2D sky images, \{$X',\lambda$\} for spectral images as
from a long-slit (or slitless) spectrometer, and \{$\lambda,t$\} for
time-resolved spectroscopy.  There is also a growing body of 3D or 4D data
sets, \{$X,Y,\lambda[,t]$\}, being produced by instruments across the
spectrum (Emsellem 2008). Here we focus on the needs of the High-Energy
Transmission Grating (HETG) spectrometer on {\it Chandra} (Canizares et al.
2005): a slitless dispersive imaging spectrometer operating in the X-ray
range of 0.3 to 8 keV, although the techniques and software (s/w) described
here are applicable to other instruments as well.

\begin{figure}[tbh]
\epsscale{0.95}
\plotone{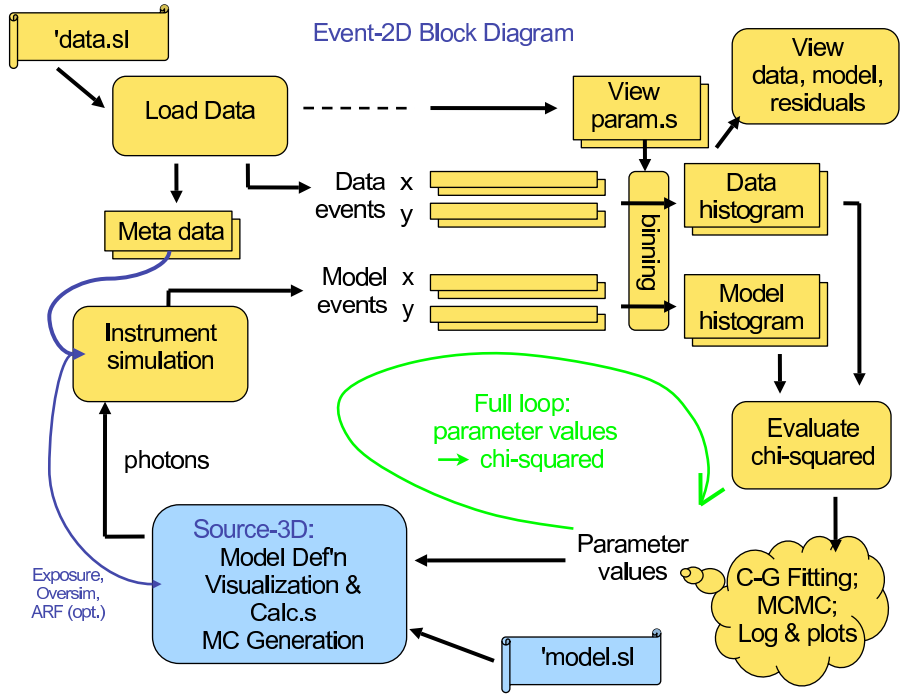}
\caption{The Event-2D system.  The flow from model parameters to a statistic
measuring data-model agreement is shown, starting at the bottom right and proceeding 
clockwise.} \label{PA.11-fig-1}
\end{figure}

For extended sources observed with the HETG, the dispersed data are 2D and
combine spatial and spectral information (Dewey 2002). A block diagram of
the Event-2D system used to analyze these data is shown in
Figure~\ref{PA.11-fig-1}.
Some key ingredients are: the ability to model the 3D geometric-spectral
properties of the source, appropriate instrument simulation
to generate modeled 2D events from the source photons, the management and
viewing of multiple 2D data sets, and a flexible and quantitative comparison
of the real and modeled data. 

Event-2D is written in S-Lang
(\htmladdURL{http://www.jedsoft.org/slang/}),
a high-performance interpreted
language which is also used in our general X-ray analysis system, ISIS
(Houck \& Denicola 2000).  S-Lang/ISIS allows us to import (interface to) a
wide variety of external modules (gsl, volpack, etc.) in order to extend
available functionality (Noble \& Nowak 2008). This HETG s/w falls under the
large umbrella of the Hydra (\htmladdURL{http://space.mit.edu/hydra/})
project at MIT.  The following sections summarize the status of the main
components of the current Event-2D system.

\section{Source Model Definition}

\begin{figure}[tbh]
\epsscale{0.95}
\plotone{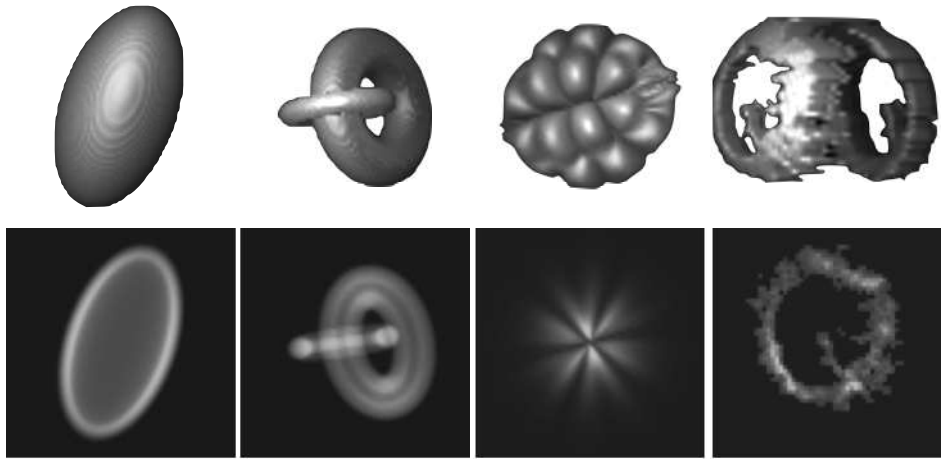}
\caption{Example shapes in the {\tt\bf v3d} library.  The upper row shows
a 3D-solid view of the objects; the lower row shows 2D projections of
their optically thin emission.} \label{PA.11-fig-2}
\end{figure}

The Source-3D s/w allows models to be created as a combination of geometric
components each having their own 3D geometry, spectral emission, and
velocity properties. Examples of some of the geometric primitives provided
by the {\tt v3d} library are shown in Figure~\ref{PA.11-fig-2}.
The {\tt v3d} routines generate 3D arrays (data cubes) of values which can be
combined, e.g., through union and intersection, to produce more complex
geometries. The emission spectra associated with each component can be
defined by the usual ISIS function specifications and ``.par'' files.
Velocity properties including ``Hubble-like'' expansion and orbital rotation
can be included in the components; these source motions are very
important for the spectral imprints they leave via Doppler shifts.  The user
specification of a source model is most conveniently done through a custom
S-Lang script file which includes definition of
the model parameters and rules for updating the model based on
the parameters. While this does require some ``programming'' on the users
part, it is relatively simple and gives complete control of the model
definition to the user: e.g., direct access to the structures that define
the spectral and the geometric components.  The source can include a foreground absorption
component which is applied in the observer's rest frame, along with an
optional instrument effective area (or first approximation thereof) allowing
the photon Monte Carlo (MC) generation and subsequent detection to be more
efficient. Besides useful model visualizations, the main output of
the model is the MC generation of ``photons'' with \{$X,Y,E,t$\} values;
these can also be used
outside of Event-2D for other applications, e.g., as input to
an observatory's simulator.

\section{Data and Instrument Simulation}

Data are loaded from {\it Chandra} FITS event files into internal structures
by specifying the 2 desired event tags,
e.g., \{SkyX,SkyY\} or \{$\lambda$,TG\_D\}.  Here also, it is convenient
to create a user script to load the data sets.
Several structures keep track of the instrument properties,
the event data, and the viewing
and binning parameters for each of the loaded data sets. Since these
structures are user accessible, custom analyses are facilitated.

X-ray instrument simulation is done by MC ray-tracing of
appropriate fidelity. The instrument knowledge is
coarse but useful: an {\tt arf} specifies
effective area vs energy, a 2-Gaussian approximates the on-axis PSF, and the
detector intrinsic energy resolution is specified by a simple equation,
$f(E)$. Grating simulation includes period variation and cross-dispersion
blurs.  The appropriate 2 simulated tags are then loaded into the model
events for comparison with the data events, Figure~\ref{PA.11-fig-1}.

\section{Data--Model Comparison and Fitting}

\begin{figure}[tbh]
\epsscale{0.95}
\plotone{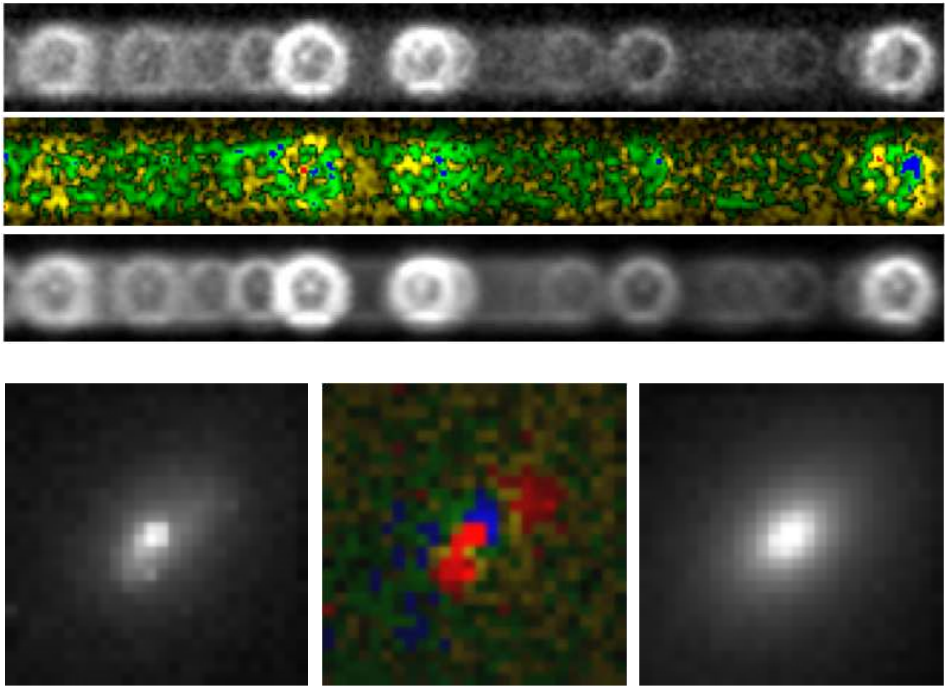}
\caption{Examples of Event-2D data. Upper panels are
HETG-dispersed spectral-images of data, residual, and model for the SNR
E0102 (top to bottom.)
Lower panels show data, residuals, and model images (l to r) for an X-ray cluster
modeled by a $\beta$-model ellipsoid plus point source.} \label{PA.11-fig-3}
\end{figure}

The data and model can be compared by
binning the events in a regular 2D grid (image) and calculating the
usual $\chi^2$ of the residuals between them as shown for two cases in
Figure~\ref{PA.11-fig-3}.
The specific feature of the model that is being fit often guides
the range and binning size for the comparison. An important aspect of the
fitting here is that the MC model itself contains ``noise'': re-evaluation
with the same input parameters does not give the identical model result.
This MC noise can be reduced below scientific relevance
by ``over simulating'', generating more events
than in the data and scaling appropriately.
Even so, the fitting has to be noise-aware and noise-tolerant.  A modified
conjugate gradient fitting method and a Markov Chain MC method are
available; each of these requires user guidance in the form of providing a "should make a
noticeable difference" size scale for each of the fitted parameters.


\acknowledgements
Support for this work was provided by NASA through the AISRP grant NNG06GE58G
and by NASA (NAS8-03060) via SAO contract SV3-73016 to MIT for support of
the {\it Chandra} X-ray Center and Instruments.

\end{document}